\documentclass[aps,prc,onecolumn,superscriptaddress,showpacs,showkeys]{revtex4}
\usepackage{graphicx}
\usepackage{bm}
\usepackage{epsfig}
\usepackage{amsfonts}
\usepackage{amssymb,amscd}
\usepackage{subfigure}
\usepackage{xcolor}

\begin{document}
\begin{flushright}
MS-TP-22-26
\end{flushright}

\title{Kaon production in high multiplicity events at the LHC}

\author{Yuri N. {\sc Lima}}
\email{limayuri.91@gmail.com}
\affiliation{Institute of Physics and Mathematics, Federal University of Pelotas, \\
  Postal Code 354,  96010-900, Pelotas, RS, Brazil}

\author{Andr\'e V. {\sc Giannini}}
\email{giannini@ifi.unicamp.br}
\affiliation{Instituto de F\'isica Gleb Wataghin, 
Universidade Estadual de Campinas, 
Rua S\'ergio Buarque de Holanda 777, 13083-859 S\~ao Paulo, Brazil}

\author{Victor P. {\sc Gon\c{c}alves}}
\email{barros@ufpel.edu.br}
\affiliation{Institut f\"ur Theoretische Physik, Westf\"alische Wilhelms-Universit\"at M\"unster,
Wilhelm-Klemm-Stra\ss e 9, D-48149 M\"unster, Germany}
\affiliation{Institute of Modern Physics, Chinese Academy of Sciences,
  Lanzhou 730000, China}
\affiliation{Institute of Physics and Mathematics, Federal University of Pelotas, \\
  Postal Code 354,  96010-900, Pelotas, RS, Brazil}

\begin{abstract}
The production of the $K_S^0$ meson in high multiplicity $pp$ collisions at $\sqrt{s} =$ 13 TeV is investigated considering the hybrid formalism and the solution of the running coupling Balitsky - Kovchegov equation. The associated cross section is estimated and compared with the experimental data for the transverse momentum spectrum. Moreover, we analyze the self-normalized yields of $K_S^0$ mesons as a function of the multiplicity of coproduced charged hadrons and demonstrate that a steep increasing is theoretically predicted. A comparison with the ALICE data is presented considering two distinct solutions of the BK equation.  
\end{abstract}
\keywords{Particle production, Color Glass Condensate Framework, Hybrid factorization formalism}
\maketitle
\date{\today}

\section{Introduction}
In recent years, different experimental collaborations at the Relativistic Heavy Ion Collider (RHIC) and at the Large Hadron Collider (LHC) have found that the $J/\Psi$, $D$, $K_S^0$ and $\Lambda$ yields observed in proton - proton ($pp$) collisions grow rapidly as a function of the multiplicities of co - produced charged particles \cite{ALICE:2015ikl,ALICE:2017wet,STAR:2018smh,ALICE:2020msa,ALICE:2020eji,ALICECollaboration:2020, ALICE:2021zkd}. Although the development of a theoretical framework for high multiplicity events has started several decades ago, the description of the current data using a unified approach remains a challenge (See e.g. Refs. \cite{Ferreiro:2012fb,Kopeliovich:2013yfa,Lang:2013ex,Werner:2013tya,Ferreiro:2015gea,Ma:2018bax,Kovner:2018azs,Weber:2018ddv, Levin:2019fvb,Kopeliovich:2019phc, Gotsman:2020ubn,Siddikov:2020lnq, Siddikov:2021cgd, Siddikov:2021dfn, Stebel:2021bbn,Salazar:2021mpv}). One the main open questions is if the modification observed in high multiplicity events compared to the minimum bias case is due to either initial or final state effects or both. The similarity of multiplicity enhancements observed in the charm and strange sectors favours the interpretation that the behaviour is due to initial state effects, but models based on very distinct underlying assumptions and physical mechanisms are also able to describe the current data. Considering the recent restart of the LHC, we expect to have in the forthcoming years a larger amount of data for the production of different hadrons in high multiplicity events, which will allow us to improve our understanding of the mechanism that generates the multiplicity enhancement.  

One of the promising frameworks to describe the particle production in low and high - multiplicity events at the LHC
is the Color Glass Condensate (CGC) formalism \cite{Gelis:2010nm}, which is an effective field theory that states that a dense system of partons is produced in hadronic collisions. Such system is characterized by a new scale, the saturation scale $Q_s$,  which increases as function of the energy, atomic number and multiplicity (See e.g. Refs. \cite{
Kovchegov:2012mbw,Morreale:2021pnn}). In this framework, high multiplicity events  are attributed to the presence of rare parton configurations (hot spots) in the hadrons that participate of the collision. Such highly occupied gluon states are characterized by larger saturation scales in comparison to the typical configurations present in minimum bias events. Naturally, in the CGC framework, low and high multiplicity approaches are expected to be described in a unified way, but with the scattering amplitude being estimated for different values of the saturation scale. {Such} assumption is the starting point of the studies performed in Refs. \cite{Levin:2019fvb,Kopeliovich:2019phc, Gotsman:2020ubn,Siddikov:2020lnq, Siddikov:2021cgd, Siddikov:2021dfn, Stebel:2021bbn,Salazar:2021mpv} and for the analysis that will be carried out in this paper, where we will focus on the production of the $K_S^0$ meson in high multiplicity events at the LHC. The strangeness enhancement has already been discussed in the literature in Ref. \cite{Siddikov:2021cgd}, where the authors have applied for the $K$ production, the dipole approach developed  for the calculation of open charm and bottom states. As a consequence, only the $g g \rightarrow s \bar{s}$ channel is taken into account and the presence of a strange (anti) quark in the wavefunction of the incident hadrons is disregarded. Moreover, these authors have assumed a phenomenological model for the description of the dipole - hadron interaction. In contrast, in this paper we will consider the hybrid formalism \cite{dhj} to treat the $K_S^0$ production and the solution of the running coupling Balitsky - Kovchegov (BK) equation to describe the evolution of the dipole - hadron scattering amplitude \cite{BAL,KOVCHEGOV,kovwei1,javier_kov,balnlo}. One has that the hybrid formalism takes into account of the contribution associated to the gluon and quark - initiated channels and, over the last decade,  has been extensively applied for the description of light particle production in hadronic colliders, with its predictions describing with reasonable success the RHIC and LHC data (See e.g. Refs. \cite{Boer:2007ug,Betemps:2008yw,Albacete:2010bs,Altinoluk:2011qy,Albacete:2012xq,Chirilli:2011km,Kang:2014lha,Altinoluk:2014eka, Iancu:2016vyg,Duraes:2015qoa,Ducloue:2017dit,Ducloue:2017mpb,Carvalho:2017zge,Liu:2019iml,Iancu:2020mos,Shi:2021hwx}). As we will demonstrate below, such approach is also able to describe the current minimum bias data for the transverse momentum spectrum of the $K_S^0$ meson, which motivates its application for high multiplicity events.

This paper is organized as follows. In the next Section, we will present a brief review of the hybrid formalism and will discuss the ingredients considered in our calculations. In Section \ref{section:results}, we will present our predictions for the transverse momentum spectrum of the $K_S^0$ meson and the contribution of the gluon and quark - initiated processes will also be estimated. Moreover, we will analyze the dependence of the $K_S^0$ and charged particle yields in the value of the saturation scale, considering distinct transverse momentum ranges and two distinct solutions of the BK equation. Predictions for the multiplicity dependence of the $K_S^0$ meson production will be compared with the experimental data from ALICE. Finally, in Section \ref{section:conc} we will summarize our main results and conclusions.

\begin{figure}[t]
		\includegraphics[scale=0.5]{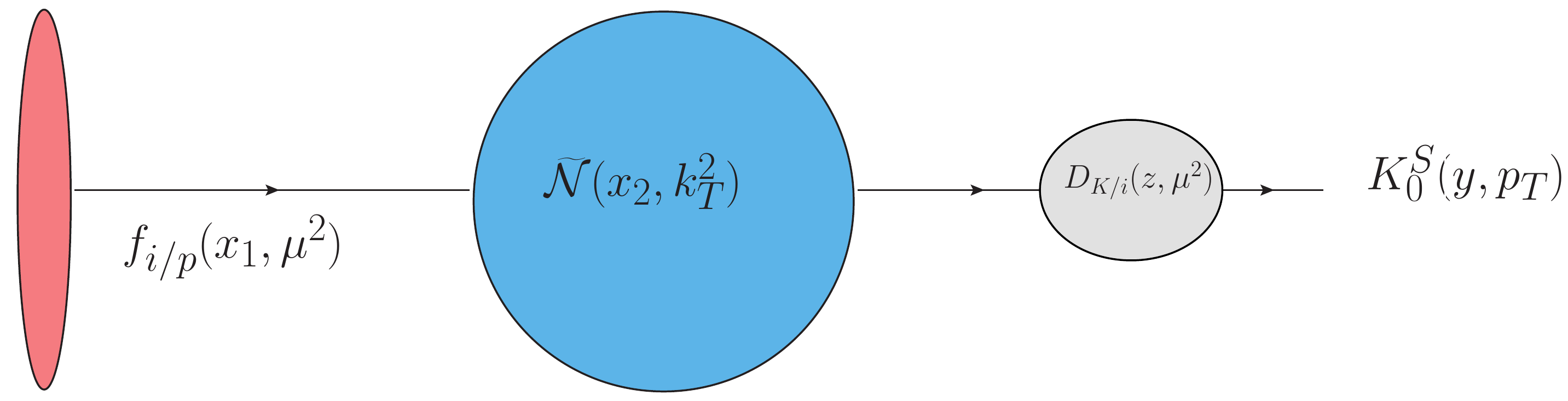}
	\caption{Representation of the $K_S^0$ meson production in the hybrid formalism. One has that a parton of the incident hadron interacts with the target and subsequently hadronizes in the $K_S^0$ meson.}
	\label{Fig:Diagram}
\end{figure}

\section{Kaon production in the hybrid formalism}
\label{section:formalism}
{ 
The treatment of the kaon production in $pp$ collisions at high energies is still a theme of debate. In principle, it can be estimated using the collinear formalism and the solutions of the DGLAP equation for the parton distribution and fragmentation functions. However, an approach based on this formalism disregards the non - linear effects in QCD dynamics that are expected to contribute at small - $x$ and high multiplicities. Such effects are taken into account by the CGC formalism, which predicts the modification of the proton wave function due to the high partonic density present when it is probed at high energies. One has that for  particle production at central rapidities, the wave functions of both projectile and target are probed for small values of $x$. The proof of the factorization theorem for this dense - dense configuration is still an open question (See e.g. Refs. \cite{Kovchegov:2001re,Gelis:2008rw,Gelis:2008ad,Gelis:2008sz,Chirilli:2015tea}). Usually, the $k_T$ - factorization is assumed to still be valid in the kinematical range probed by LHC, and the particle production at midrapidities is estimated using  the unintegrated gluon distributions (UGDs) as input, which are obtained by solving the Balitsky - Kovchegov equation at different levels of sophistication (See e.g. Refs. \cite{Duraes:2016yyg,Dumitru:2018gjm,Zhao:2022ubw}). Some of the shortcomings of this formalism are that the contribution of the quark - initiated processes are disregarded and it is not clear the rapidity range in which its predictions are valid.  On the other hand, for  particle production at forward rapidities, one has that the wave function of one of the projectiles is probed at large Bjorken $x$ and that of the other at very small $x$. In this dilute - dense configuration, the cross sections can be estimated using hybrid formalism. Over the last decade, the hybrid formalism has been developed, improved and  applied for the description of hadron production in hadronic collisions at RHIC and LHC \cite{
Boer:2007ug,Betemps:2008yw,Albacete:2010bs,Altinoluk:2011qy,Albacete:2012xq,Chirilli:2011km,Kang:2014lha,Altinoluk:2014eka, Iancu:2016vyg, Duraes:2015qoa,Ducloue:2017dit,Ducloue:2017mpb,Carvalho:2017zge,Liu:2019iml,Iancu:2020mos,Shi:2021hwx}. The basic assumption is that the cross section for particle production can be  expressed as a convolution of the standard parton distributions, the  scattering amplitude (which includes the high-density
effects) and the parton fragmentation functions. One of the advantages of this formalism is that it takes into account the quark and gluon initiated processes. However, the contributions of the non - linear effects for the projectile are disregarded, which implies that the validity range of its predictions for $y \rightarrow 0$ is not yet well determined.  
As already emphasized in Ref.  \cite{Albacete:2012xq}, the corresponding limits of applicability of the $k_T$ - factorization and hybrid formalisms are not clear and have only been estimated on an empirical basis.

Another important aspect is that the above considerations are valid for the typical color charge configurations in the proton wave functions, which dominate the description of minimum bias collisions. For rare high multiplicity events, one expects that the  projectile and/or target wave functions will be characterized by a saturation scale larger than the average one. Therefore, even for central rapidities, a high multiplicity event can be generated by a collision between two asymmetric systems, i.e. by a dilute - dense configuration. Such aspects justify a phenomenological analysis of the high multiplicity events using the hybrid formalism. In what follows, we will demonstrate that this approach is able to describe the current data for the $K_S^0$ spectrum, providing  similar results to those derived using the $k_T$ - formalism. Moreover, we will compare our predictions with the current data for high multiplicity events, measured for midrapidities, and results for forward rapidities will also be provided. 
}

The representation of the hybrid formalism for the $K_S^0$ meson production is presented in Fig. \ref{Fig:Diagram}.  Taking into account of the gluon and quark - initiated subprocesses, one has that the invariant yield for single-inclusive $K_S^0$ production in hadron-hadron processes will be described in the CGC formalism as follows~\cite{dhj}
\begin{eqnarray}
{dN_{K_S^0} \over dy d^2p_T} &=& 
{{\cal{K}} \over (2\pi)^2} \int_{x_F}^{1} dx_1 \, {x_1\over x_F}
\Bigg[\sum_{q = u,d,s} f_{q/p}(x_1,\mu^2)\, \widetilde{\cal N}_{F} \left(k_T,x_2\right)\,
D_{K_S^0/q}\left({x_F\over x_1},\mu^2\right)
\nonumber \\
& &+~
f_{g/p}(x_1,\mu^2)\, \widetilde{\cal N}_{A} \left(k_T,x_2\right)\, 
D_{K_S^0/g}\left({x_F\over x_1},\mu^2\right)\Bigg]~
\label{Eq:hybrid}\,.
\end{eqnarray}
One has that $p_T$, $y$ and $x_F$ are the transverse momentum, rapidity and the Feynman-$x$ of the produced hadron, respectively. Moreover, $x_1$ denotes the momentum fraction of a projectile parton, $x_F=\frac{p_T}{\sqrt{s}}e^{y}$ and $k_T = {x_1\over x_F}p_T$, with  the momentum fraction of the target parton 
being given by $x_2=x_1e^{-2y}$. One also has that
 $f_{i/p}(x_1,\mu^2)$ are the projectile parton
distribution functions and $D_{K_S^0/i}(z, \mu^2)$ are the parton fragmentation functions into the $K_S^0$ meson. As in previous studies \cite{Boer:2007ug,Betemps:2008yw,Albacete:2010bs,Altinoluk:2011qy,Albacete:2012xq,Chirilli:2011km,Kang:2014lha,Altinoluk:2014eka, Iancu:2016vyg,Duraes:2015qoa,Ducloue:2017dit,Ducloue:2017mpb,Carvalho:2017zge,Liu:2019iml,Iancu:2020mos,Shi:2021hwx}, we will assume that the parton distribution and fragmentation functions  evolve according to the DGLAP evolution equations \cite{dglap} and obey the momentum
sum-rule. In particular, we will consider the CT14~\cite{ct14} and AKK08~\cite{Albino:2008fy} parametrizations for these quantities. The factorization scale $\mu^2$ will be assumed 
as being $\mu^2 = {\rm max }(Q_s^2,p_T^2)$. Previous results derived using the hybrid formalism have pointed out that in order to describe the data for different kinematical ranges one has to adjust the normalization by a  ${\cal{K}}$-factor, which can be energy and rapidity dependent. Such factor is interpreted as taking account of higher order corrections and/or of other dynamical 
effects not included in the CGC formulation. Such uncertainty will not affect the analysis of high multiplicity events, since this factor is expected to be same in low and high multiplicity events and, consequently, it cancels in our prediction for the ratio between the results for high multiplicity and minimum bias events.

One has that the predictions for the $K_s^0$ yield are also strongly dependent on the description of the fundamental and adjoint representations of the forward dipole amplitude, $\widetilde{{\cal{N}}}_F$  and  $\widetilde{{\cal{N}}}_A$, respectively. In the CGC framework, the forward dipole scattering amplitudes encodes all the information about the hadronic scattering, and thus about the non-linear and quantum effects in the hadron wave function.  Such quantities can be expressed either in the momentum  or in the position spaces, with the representations related by 
\begin{eqnarray}
\widetilde{{\cal{N}}}_{A,F}(x,p_T)=  \int d^2 r \,  e^{i\vec{p_T}\cdot \vec{r}}\left[1-{\cal{N}_{A,F}}(x,r)\right]\,\,,
\end{eqnarray}
In our analysis we will consider the solutions of the running coupling Balitsky–Kovchegov equation for ${\cal{N}_{A,F}}(x,r)$ derived in Ref. ~\cite{Albacete:2012xq} considering two distinct initial conditions characterized by 
an anomalous dimension larger than unity. 

Following previous studies of the high multiplicity events using the CGC formalism, we will assume that the particle production mechanism is the same for low and high - multiplicity events, with the main difference being the saturation scale present in these two classes of events. In other words, we will assume that Eq. (\ref{Eq:hybrid}) is valid for both classes, and that the high multiplicity configurations can be approximated by increasing the value of $Q_s$ as follows $Q_s^2(x,n) = n \cdot Q_s^2(x)$, where $n$ characterizes the multiplicity. Such quantity is given approximately by  the multiplicity of charged particles weighted by their minimum bias value: $n \approx dN_{ch}/d\eta/ \langle dN_{ch}/d\eta \rangle$. As in Ref. \cite{Ma:2018bax},  we will solve the BK equation considering multiples of initial saturation scale at $x = 0.01$,  $Q_0^2 = 0.168$ GeV$^2$, which has been determined from fits to the minimum bias $ep$ HERA data. Such modification naturally implies larger values for the saturation scale probed in a given event, which depends on $x$ and has its evolution determined by the BK equation.  

\begin{figure}[t]
	\centering
	\subfigure[]{
		\includegraphics[scale=0.68]{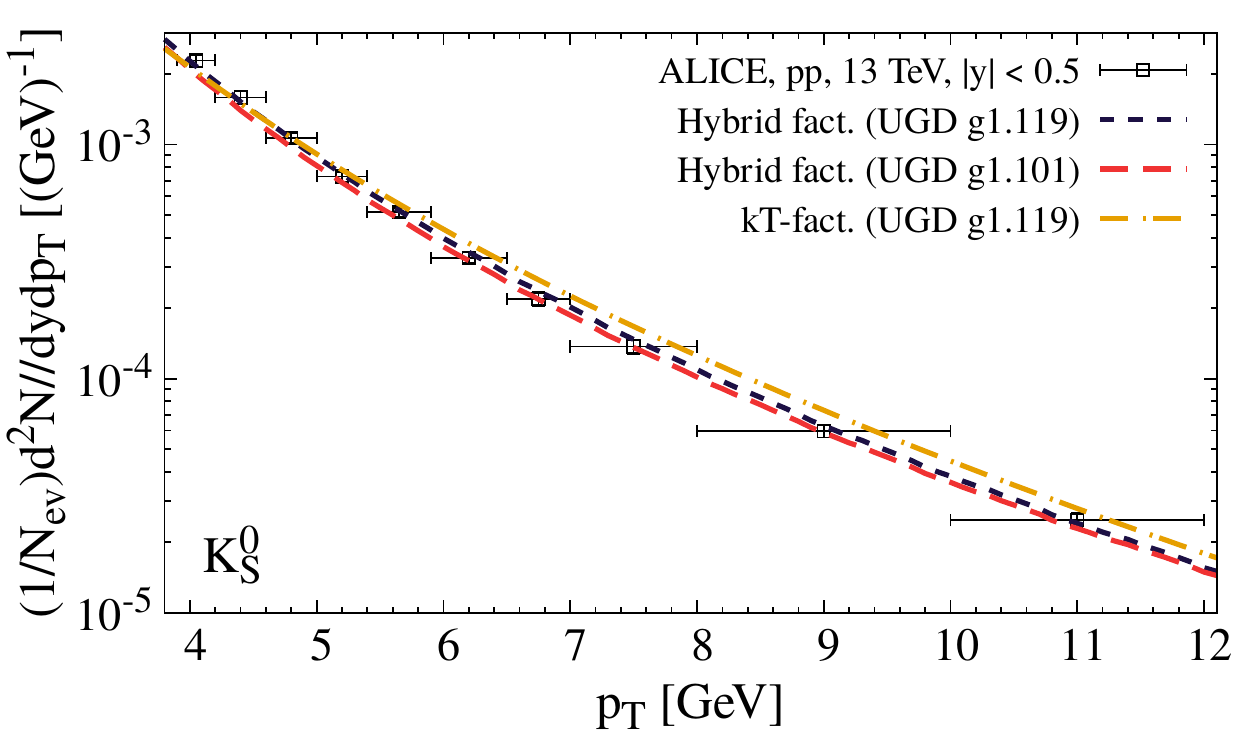}}
	\centering
	\subfigure[]{
		\includegraphics[scale=0.68]{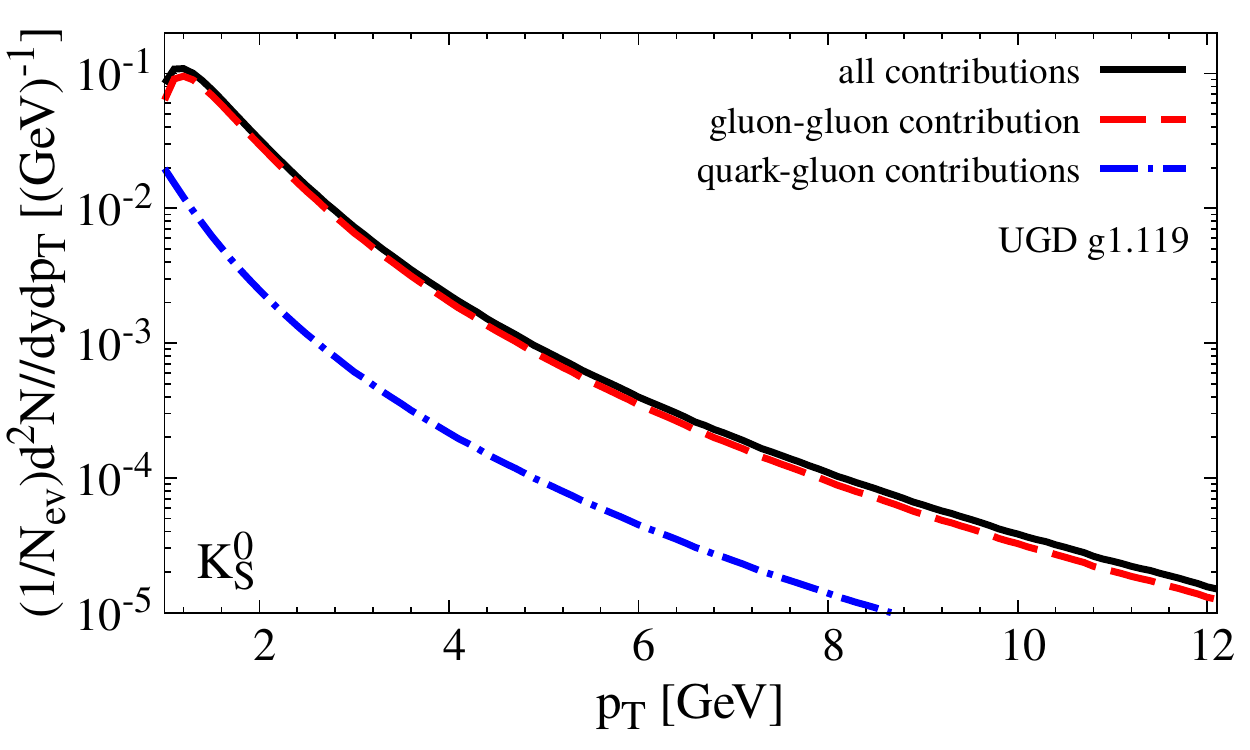}}
	\caption{(a) Dependence of the self-normalized yield for the $K^0_S$-meson production on the transverse momentum ($p_T$)  compared to experimental data provided by the ALICE collaboration \cite{ALICECollaboration:2020}. (b) Contribution of the gluon and quark - initiated subprocesses for the $K^0_S$-meson production derived using the  UGD g1.119 model.}
	\label{Fig:spectra}
\end{figure}

\section{Results and discussion}
\label{section:results}

Initially, let's compare our predictions for the transverse momentum distribution of the $K^0_S$ meson with the experimental data from the ALICE collaboration \cite{ALICECollaboration:2020} for the self normalized yields $(1/N_{ev})d^2N/dydp_T$. We will consider the solutions of the BK equation derived assuming two distinct initial conditions~\cite{Albacete:2012xq}, which will be denoted by ``UGD g1.101'' and ``UGD g1.119'' hereafter. As already emphasized in Ref. \cite{Siddikov:2021cgd}, the prediction of the normalization of the self normalized yields is strongly affected by non-perturbative effects, since this quantity is dominated by the contribution of kaon production at very low $p_T$, which cannot be evaluated reliably in our approach. As a consequence, in our analysis we will adjust the normalization of our predictions in order to describe the data.  In Fig. \ref{Fig:spectra}(a) we present our results.  { For comparison we also present the spectrum derived using the $k_T$ - factorization formalism and the UGD g1.119. One has that both the hybrid and the $k_T$ - factorization  formalisms describe the current date for the $K_S^0$ production in $pp$ collisions. In what follows, we will restrict our analysis to the hybrid formalism. Regarding the predictions derived using this formalism and distinct models for the unintegrated gluon distributions, one has that the shape of the spectrum  can be quite well described by both UGDs considered.}
The contribution of the gluon and quark - initiated channels for the $K_S^0$ production is presented separately in Fig. \ref{Fig:spectra}(b) for the UGD g1.119 model. One has verified that similar results are obtained for the g1.101 model. Our results indicate that the gluon channel is dominant in the kinematical range considered, which can explain why the approach used in Ref. \cite{Siddikov:2021cgd} is also able to describe the ALICE data.

\begin{figure}[t]
	\centering
	\subfigure[]{
		\includegraphics[scale=0.68]{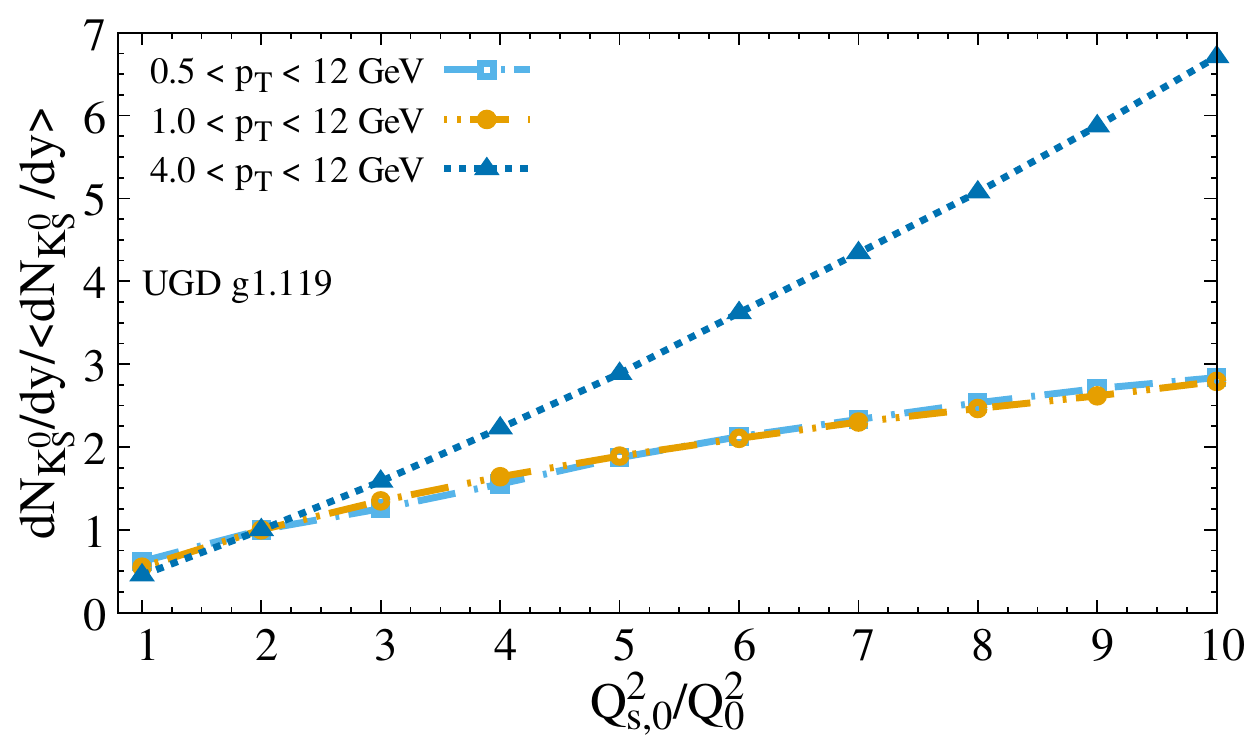}}
	\centering
	\subfigure[]{
		\includegraphics[scale=0.68]{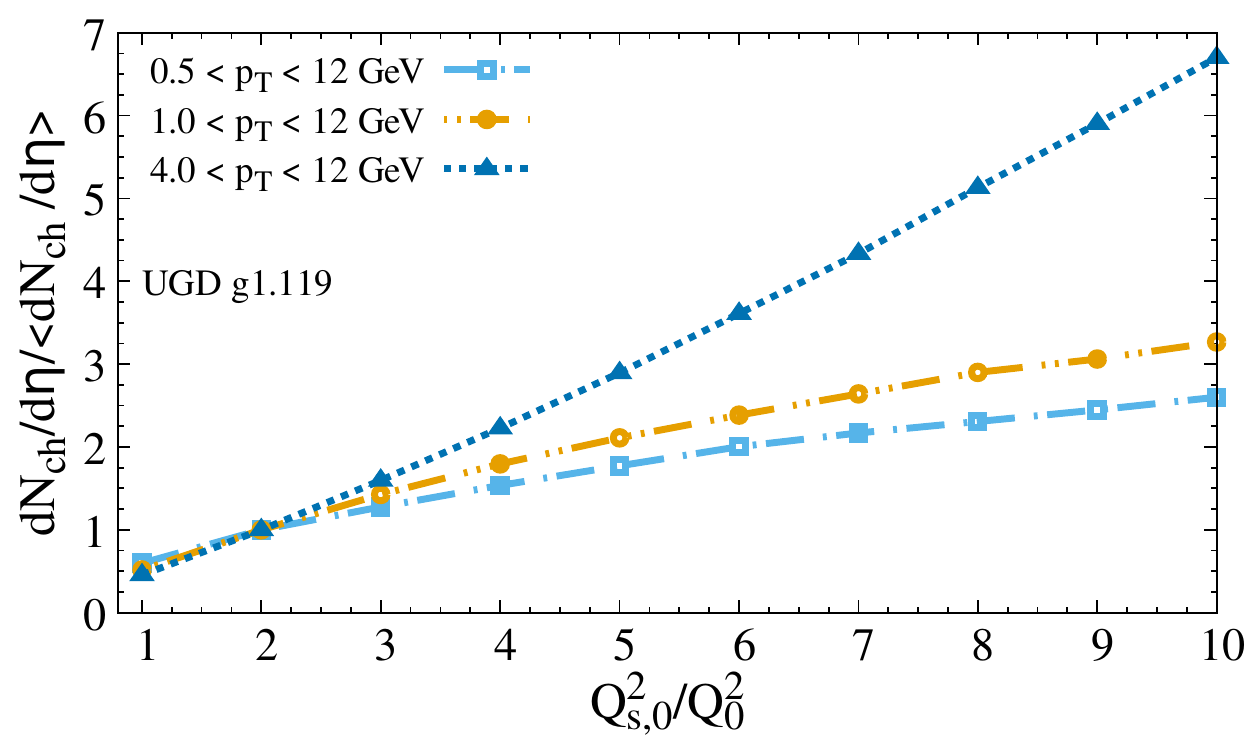}} \\
	\subfigure[]{	\includegraphics[scale=0.68]{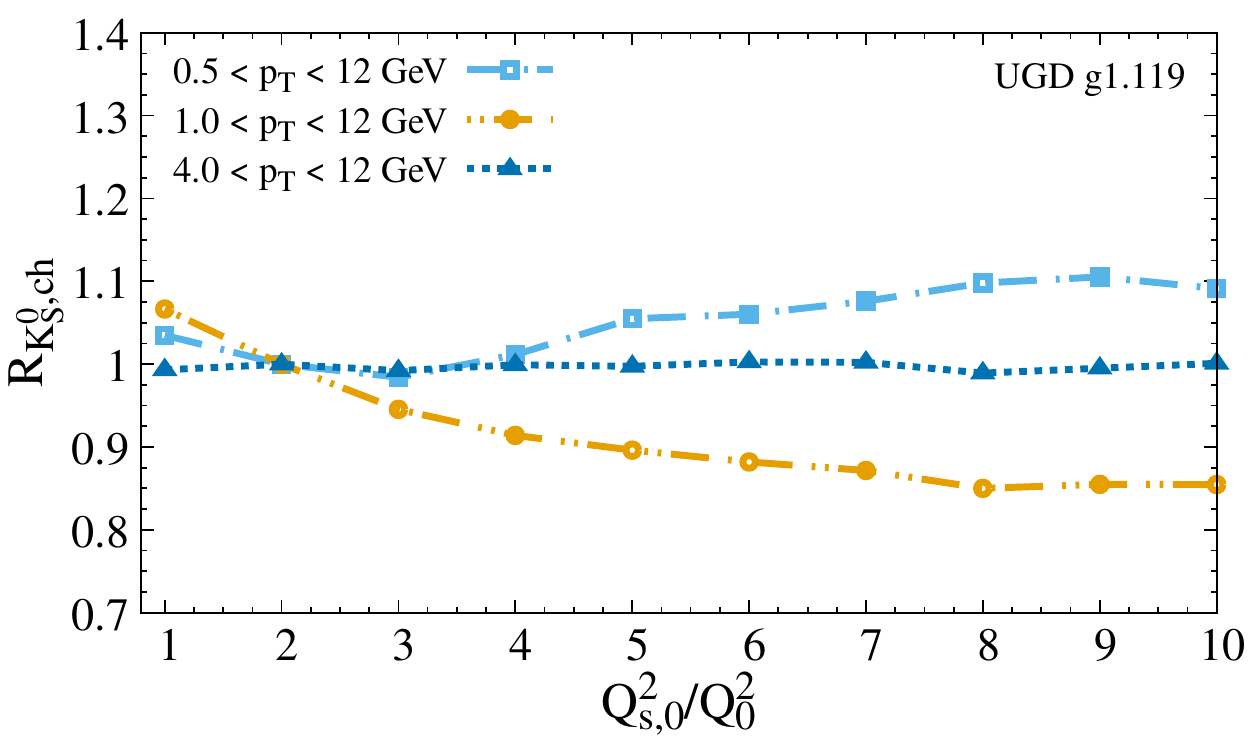}}
	\caption{Relative multiplicity of (a) $K_S^0$ mesons and (b) charged particles as a function of $Q^2_{s,0}/Q^2_0$, considering different ranges of the transverse momentum $p_T$. (c) Ratio between the predictions for the production of $K_S^0$ mesons and  charged particles as a function of $Q^2_{s,0}/Q^2_0$. Results derived using the  UGD g1.119.}
	\label{Fig:multiplicity-Q2S_charged}
\end{figure}

In order to estimate the impact of varying the initial saturation scale, in Figs. \ref{Fig:multiplicity-Q2S_charged} (a) and (b) we present our results for the relative multiplicity of $K_S^0$ mesons and charged hadrons, respectively, for a $pp$ collision at 13 TeV as a function of  $Q^2_{s,0}/Q^2_0$,  with $Q_0^2 = 0.168$ GeV$^2$. We present our predictions, derived assuming the UGD g1.119 model, for different range of the transverse momentum $p_T$. For charged hadrons, we also have estimated the yield using the hybrid formalism, considering the contribution of charged pions, baryons  and  strange mesons. It is important to emphasize that we have verified that the corresponding predictions describe the current data for the inclusive hadron production at central rapidities in $pp$ collisions at the LHC. One has that the predictions for $K_S^0$ mesons and charged hadrons are similar, but the increasing with $Q^2_{s,0}/Q^2_0$  is  dependent of the $p_T$ range considered. Such result is expected since the impact of the nonlinear effects are  strongly dependent if $Q_s^2$ is larger or smaller than $p_T^2$. Events where $Q_s^2 \gtrsim p_T^2$ are expected to be determined by the nonlinear QCD dynamics.  Therefore, a larger value of the minimum value of $p_T$, implies a reduction of the number of events produced in the saturated regime. The differences between the predictions for $K_S^0$ mesons and charged hadrons can be quantified by calculating the ratio between the corresponding results, 
\begin{equation}
R_{K_S^0,ch}=\frac{dN_{K_S^0}/dy/\langle dN_{K_S^0}/dy\rangle}{dN_{ch}/d\eta/\langle dN_{ch}/d\eta \rangle},
\end{equation}
with the results being presented in Fig. \ref{Fig:multiplicity-Q2S_charged} (c).
One has that when only large $p_T$ values are considered, the ratio is consistent with unity, indicating that the production of $K_S^0$ mesons and charged hadrons are similarly affected by the saturation effects. On the other hand, if events with lower transverse momentum are included, one has that the behavior with $Q^2_{s,0}/Q^2_0$ becomes final state dependent.

\begin{figure}[t]
	\centering
	\subfigure[]{
		\includegraphics[scale=0.68]{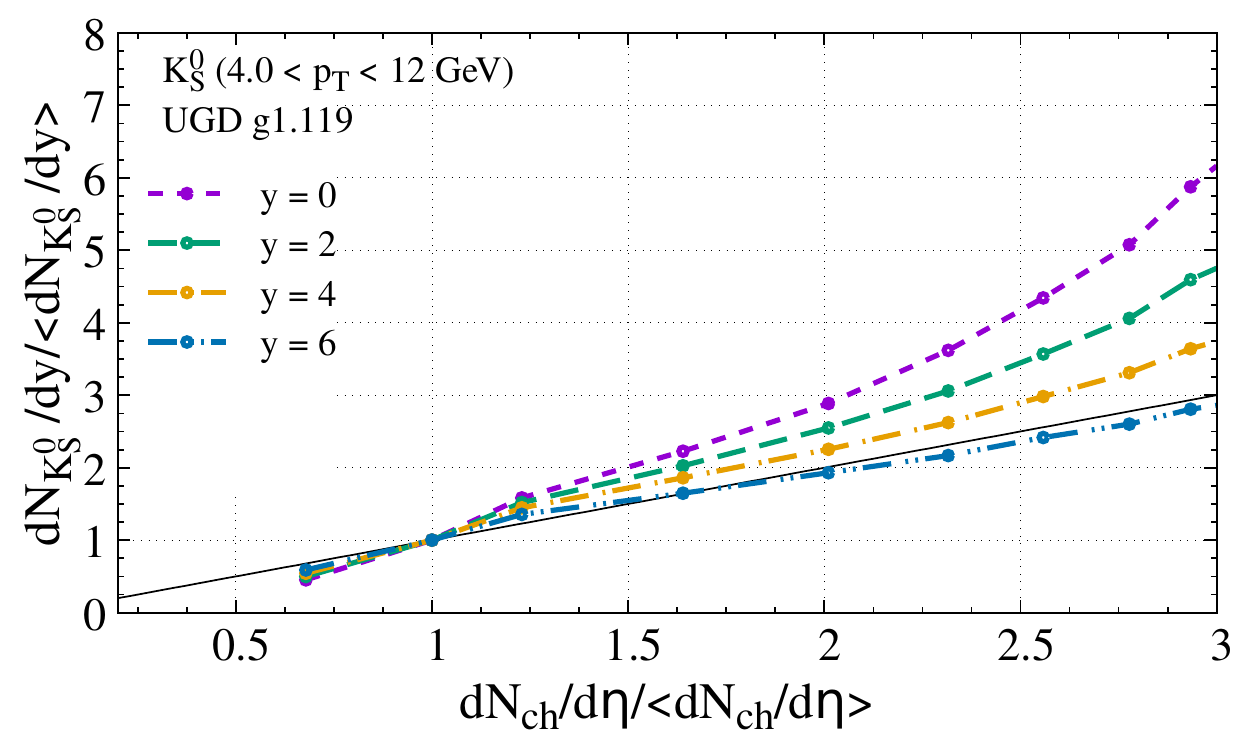}}
	\centering
	\subfigure[]{
		\includegraphics[scale=0.68]{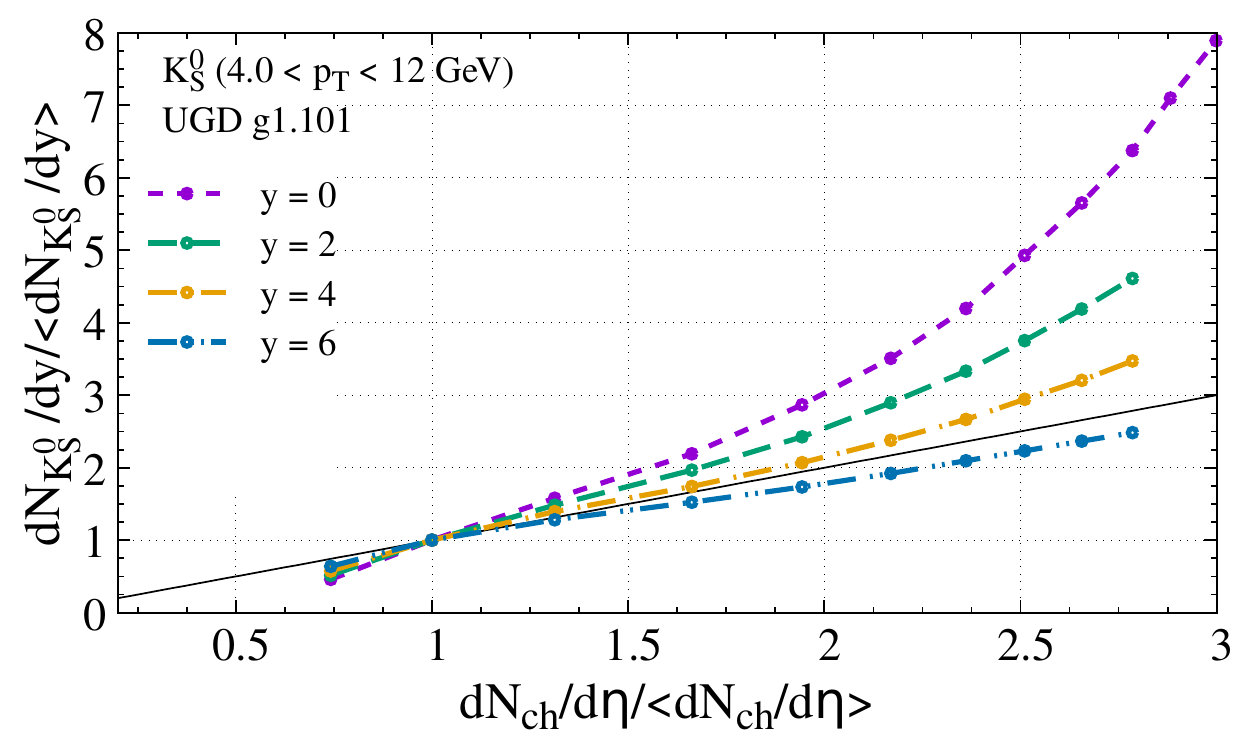}} 	\caption{Correlation between the normalized $K_S^0$ and charged hadron yields in $pp$ collisions at $\sqrt{s} =$13 TeV for different rapidities and considering two distinct solutions of the BK equation: (a) UGD g1.119 and (b) UGD g1.101. The solid line indicates the expected result for a linear correlation between the yields.}
	\label{multiplicitydepedence_rapidities}
\end{figure}

{
In Fig. \ref{multiplicitydepedence_rapidities} we present our predictions for the multiplicity dependence of the  
normalized $K_S^0$ yield in $pp$ collisions at 13 TeV for different rapidities, a fixed $p_T$ range and assuming two distinct solutions of the BK equation. As in Ref. \cite{ALICE:2021zkd}, the charged hadron yield will be estimated in all cases assuming that the particles are produced at central rapidities.  One has that the increaseof the $K_S^0$ yield  with the multiplicity  is strongly dependent on the rapidity. In particular, for very forward rapidities, we predict an almost linear dependence of the multiplicity, with the results derived using the two models for the UGDs being similar. Such result is expected, since for large values of $y$ one has large values of the saturation scale, implying that both the $K_S^0$ and charged hadron yields will be impacted by the non - linear effects in the QCD dynamics in a similar way. 
}

\begin{figure}[t]
	\centering
	\includegraphics[scale=0.9]{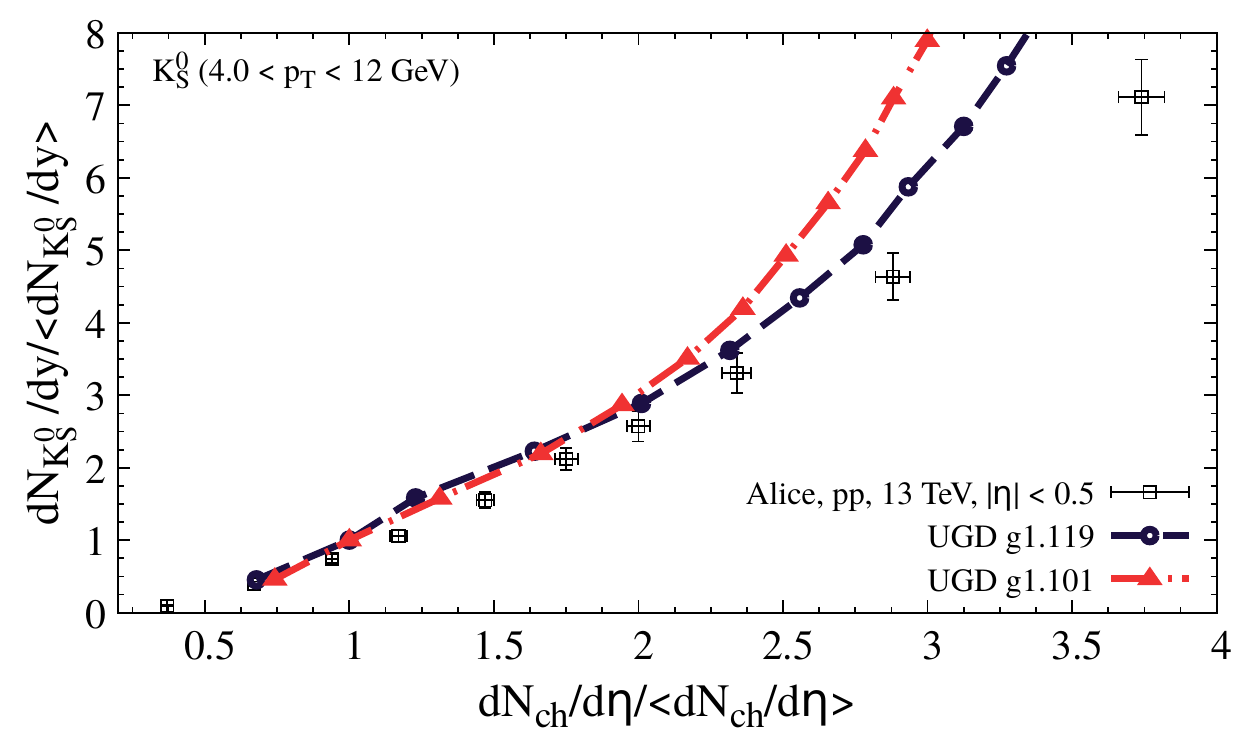}
	\caption{Correlation between the normalized $K_S^0$ and charged hadron yields in $pp$ collisions at 13 TeV derived considering two distinct solutions of the BK equation. Data from  the ALICE Collaboration \cite{ALICECollaboration:2020}.}
	\label{Fig:multiplicitydepedence}
\end{figure}

In Fig. \ref{Fig:multiplicitydepedence} we present a comparison between our predictions for the multiplicity dependence of $K_S^0$ mesons and the experimental data from the ALICE Collaboration \cite{ALICECollaboration:2020}. It is important to emphasize that these data were collected at central rapidities $|\eta| \le 0.5$ and for transverse momentum bins in the $4.0 < p_T < 12$ GeV range. We present our predictions for the two initial conditions of the BK equation considered in our analysis. Our results indicate that the corresponding predictions are similar for low multiplicities, $dN_{ch}/d\eta/ \langle dN_{ch}/d\eta \rangle \le 2.0$, but differ for larger multiplicities. One has that the UGD g1.119 model provides a better description of the data, but also fails in the description of the data for the largest value of the multiplicity. { As  $\langle dN_{ch}/d\eta \rangle \approx 15$ \cite{ALICECollaboration:2020} in the region where our description starts to overshoot the data, one has that these results indicate that the hybrid approach considered in our analysis needs to be supplemented by other ingredients when the multiplicity becomes larger than 30}. A similar overshooting is also observed in Ref. \cite{Ma:2018bax} for the $D$ meson production when events with larger transverse momentum are considered and in Ref. \cite{Siddikov:2021cgd}  for the $K_S^0$ meson production. One possible interpretation of this overestimation in the hybrid approach is that  for  high multiplicities, saturation effects cannot be neglected in the projectile. Another possible effect, not considered in our analysis, is the modification of the fragmentation function in high multiplicity events. Our results indicate that such possibilities should be analyzed in the future, which we intend to perform in a forthcoming study. { In particular, we intend to investigate pion production in high multiplicity events, which was not yet studied in the literature. Such analysis is motivated by the fact that the results presented in Refs. \cite{Goncalves:2012bn,Deng:2014vda,Albacete:2016tjq} 
indicate that the hybrid formalism is able to describe the minimum bias events at forward rapidities, but new ingredients are needed to describe the LHCf data at ultraforward rapidities ($y > 8.0$). The inclusion of these ingredients on the treatment of high multiplicity events can be useful to improve our understanding of QCD dynamics when the saturation scale is much larger than the typical value  probed at central rapidities and minimum bias events.}

\section{Conclusions}
\label{section:conc}
The description of high multiplicity events observed in small collision systems still remains an open question. In recent years, new experimental data have renewed the interest in the topic, strongly motivating a large phenomenology based on models that take into account either initial or final state effects or both. In this paper we have focused on initial state effects, as described by the the CGC framework, and applied the hybrid formalism for the description of the $K_S^0$ production in high multiplicity events present in $pp$ collisions at $\sqrt{s} = 13$ TeV. Such formalism has been successfully applied for the description of the particle production in hadronic colliders and takes into account of the gluon and quark - initiated subprocesses. Moreover, its predictions can be derived using the solutions of the running coupling BK equation. It has been demonstrated that this formalism is also able to describe the current data for the transverse momentum spectrum and that the $K_S^0$ production is dominated by the gluon - initiated subprocess. Following the study performed in Ref. \cite{Ma:2018bax}, which is able to describe the $D$ meson and $J/\Psi$ production at high multiplicities, we derived our predictions assuming that high multiplicity configurations can be approximated by increasing the value of the initial saturation scale in the BK evolution. The dependence of the relative multiplicity of $K_S^0$ mesons and charged particles with the initial condition for the BK evolution equation has been studied and we verified that they are slightly different, depending of the $p_T$ range considered.
Finally, we have compared our predictions with the ALICE data for the multiplicity dependence of $K_S^0$ mesons. Our results indicate that the hybrid formalism can describe the current data for values of multiplicity smaller than 2.5, but overestimate the data for larger multiplicities, which indicate that other effects and/or higher order corrections should be taken into account in this kinematical range. Surely, more data for the production of strange and charmed mesons in very high multiplicity events will be very useful to improve our understanding of the dynamical effects present in the new kinematical range.

\section*{Acknowledgements}
This work was partially supported by INCT-FNA (Process No. 464898/2014-5).
V.P.G. was partially supported by the CAS President's International Fellowship Initiative (Grant No.  2021VMA0019) and by CNPq, CAPES and FAPERGS. Y.N.L. was partially financed by CAPES (process 001). 
The work of A.V.G. has been supported by FAPESP through grants 17/05685-2 and 21/04924-9.

\end{document}